\documentclass[superscriptaddress,preprint,showpacs,preprintnumbers,amsmath,amssymb,aps]{revtex4}

\usepackage{graphics}
\usepackage{graphicx}
\usepackage{epsfig}
\usepackage{amssymb}
\usepackage{color}
\usepackage{dcolumn}%
\usepackage{bm}

\begin{document}

\title{Adaptive simplification of complex multiscale systems}

\author{Eliodoro Chiavazzo}\email{eliodoro.chiavazzo@polito.it}
\affiliation{Department of Energetics, Politecnico di Torino,
10129 Torino, Italy}

\author{Ilya Karlin}\email{karlin@lav.mavt.ethz.ch}
\affiliation{Aerothermochemistry and Combustion Systems Lab, ETH Zurich, 8092 Zurich, Switzerland}
\affiliation {School of Engineering Sciences, University of Southampton, SO17 1BJ Southampton, UK}

\date{\today}

\begin{abstract}
A fully adaptive methodology is developed for reducing the complexity of large dissipative systems. This represents a significant step towards extracting essential physical knowledge from complex systems, by addressing the challenging problem of a minimal number of variables needed to exactly capture the system dynamics. Accurate reduced description is achieved, by construction of a hierarchy of slow invariant manifolds, with an embarrassingly simple implementation in any dimension. The method is validated with the auto-ignition of the hydrogen-air mixture where a reduction to a cascade of slow invariant manifolds is observed.
\end{abstract}


\maketitle

\section{Introduction}\label{intro.sec}
Detailed reaction mechanisms typically serve as accurate models of dissipative complex systems with many interacting components: Biochemical processes in living cells and combustion phenomena are prototypical examples of such systems \cite{Jamshidi2008,Endy01,Limech}.
Modern research has to cope with an increasing complexity mainly in two aspects: First, the number of degrees of freedom (scaling with the number of components) is tremendously large; second, complex system dynamics is characterized by a wide range of time-scales.
For example, 
the usage of detailed reaction mechanisms in the reactive flow simulation soon becomes intractable even for supercomputers, particularly in the turbulent combustion of even "simplest" fuels such as hydrogen \cite{NZZ,Pizza2010,GoussisMaas}.
As a result, there is a strong demand for methodologies capable of both drastically reducing the description of complex systems with a large number of variables, and concurrently allowing physical insights to be gained.
Modern automated approaches to model reduction are based on the notion of {\em low dimensional manifold} of the slow motions (slow invariant manifold - SIM - for short) in the phase-space describing the asymptotic system behavior. Although several methodologies have been suggested in the literature \cite{ModRedCollection}, the construction of accurate reduced description remains a rather challenging task. 
%
In particular, the evaluation of numerical SIM approximations in the phase-space is hindered by several difficulties as far as the choice of the manifold dimension is concerned, since the latter information is typically not known {\em a priori}. In addition, accurate simplification of complex multiscale systems often requires the construction of {\em heterogeneous} (variable dimension) manifolds with the dimension $d$ ranging from unity up to tens in different regions of the phase-space. To the best of our knowledge, at the present, fully adaptive model reduction methodologies capable to cope with the above issues are still missing. This research area is pretty active and much effort has been devoted to devising techniques with the above features. The {\em intrinsic low dimensional manifold}
(ILDM) approach \cite{MP92}, the {\em computational singular perturbation} (CSP) method \cite{LG94} and the {\em minimal entropy production trajectory} (MEPT) method \cite{Leb08}
are only some representative examples.
In addition, the minimal number of reduced degrees of freedom underling the asymptotic dynamics of complex multiscale systems is still a debated issue \cite{OttingerPRL}. In this respect, we notice that, though here we mainly focus on chemical kinetics, our results have direct implications on the study of the {\em homogeneous isotropic Boltzmann equation} which has been stated a fundamental problem of Physics \cite{Truesdellbook}. The latter investigation is beyond the scope of this work, however future works shall move in this directions, where we can take advantage of recently introduced models such as the one proposed in \cite{Asinari2010}.

In the present work, we introduce a methodology which enables to cope with the accurate reduced description of large dissipative systems, where no {\em a priori} assumptions on the least number of fundamental (slow) variables are made. Toward this end, both global and local construction of slow invariant manifolds, with an embarrassingly simple implementation up to any dimension, is worked out. 

This paper is organized in sections as follows. In the section \ref{reaction.kinetics.sec}, we briefly review the governing equations for chemical kinetics. The problem of model reduction, as understood by the Method of Invariant Manifold (MIM), is discussed in the section \ref{film.equation.sec}. The Relaxation Redistribution Method (RRM) is introduced in the section \ref{RRM.gen.sec}, where both a global (section \ref{RRM.global.sec}) and a local (section \ref{RRM.global.sec}) formulation are presented. The latter methodology is validated for a detailed chemical kinetics describing a reacting mixture of hydrogen and air in the section \ref{illustration.h2o2}. Finally, conclusions are drawn in the section \ref{conclusion.sec}.

\section{Dissipative reaction kinetics}\label{reaction.kinetics.sec}
%
In the present study, we assume that a complex dissipative dynamics is governed by an autonomous system in terms of the state $\psi$ on a phase space $U$ with a unique steady state,
\begin{equation}\label{kin_sys}
\frac{{d\psi }}{{dt}} = f \left( \psi  \right).
\end{equation}
Important example of (\ref{kin_sys}) to be addressed below is the reaction kinetics where $\psi=\left( {\psi _1 , \ldots ,\psi _n } \right)$ is a $n$-dimensional vector of concentrations of various species, while the vector field $f$ is constructed according to a detailed reaction mechanism as described below.
More specifically, in a closed reactive system, the complex reaction of $n$ chemical species $A_1,...,A_n$ and $d$ elements can be represented by a (typically) large number $r$ of elementary steps: 
\begin{equation}\label{elementary.step}
\sum\limits_{i = 1}^n {\alpha _{si} A_i }  \mathbin{\lower.3ex\hbox{$\buildrel\textstyle\rightarrow\over
{\smash{\leftarrow}\vphantom{_{\vbox to.5ex{\vss}}}}$}} \sum\limits_{i = 1}^n {\beta _{si} A_i } ,\quad s = 1,...,r,
\end{equation}
where $\alpha_{si}$ and $\beta_{si}$ are the stoichiometric coefficients. The latter coefficients enable to define the three stoichiometric vectors: $\alpha_s=(\alpha_{s1},...,\alpha_{sn})$, $\beta_s=(\beta_{s1},...,\beta_{sn})$ and $\gamma_s=\beta_s-\alpha_s$, where the index $s$ runs over the $r$ elementary reactions (\ref{elementary.step}). For clarity, in the detailed reaction mechanism for air and hydrogen to be considered below \cite{Limech}, $s$ identifies any of the 21 reactions in Table \ref{table.mech}, while the corresponding stoichiometric coefficients  $\alpha_{si}$ and $\beta_{si}$ indicate the number of molecules of species $i$ in the reactants and products of reaction $s$, respectively. Production (or depletion) rates of chemical species can be conveniently expressed in terms of the differences: $\gamma_{si}=\beta_{si}-\alpha_{si}$.

Expressing the state in terms of the molar concentrations
$\psi=(c_1,...,c_n)$ (ratios of the number of moles by the volume), all chemical species evolve in time according to the mechanism (\ref{elementary.step}):
\begin{equation}\label{mol.concentration.eq}
\frac{{d \psi}}{{dt}} = \sum\limits_{s = 1}^r {\gamma _s } W_s \left( \psi, \theta \right),
\end{equation}
where $W_s\left( \psi \right)$ is the reaction rate function of the reaction $s$, which (usually) takes a polynomial form according to the {\em mass action law}:
\begin{equation}\label{mass.action.law}
W_s \left( \psi \right) = W_s^ +  \left( \psi, \theta \right) - W_s^ -  \left( \psi, \theta \right) = k_s^ + \left( \theta \right) \prod\limits_{i = 1}^n {c_i^{\alpha _i }  - k_s^ -  \left( \theta \right) \prod\limits_{i = 1}^n {c_i^{\beta _i } } } ,
\end{equation}
with the reaction constants $k_s^+$ and $k_s^-$ depending on the system temperature $\theta$ according to the Arrhenius equation:
\begin{equation}\label{arrhenius}
k_s \left( \theta \right)=A_s \theta^{n_s} e^{-E_{as} / \mathcal{R} \theta},
\end{equation}
where the quantities $A_s$, $n_s$, $E_{as}$ are fixed (and tabulated, see e.g. Table \ref{table.mech}) and referred to as pre-exponential factor, temperature exponent, activation energy of the reaction $s$, respectively, while $\mathcal{R} $ is the universal gas constant.
Due to the {\em principle of detailed balance}, a relationship between the latter reaction constants ($k_s^+$, $k_s^-$) is established for each step $s$ at the steady state: $W_s^+=W_s^-$. In general, the system (\ref{mol.concentration.eq}) is to be solved in combination with an additional equation ruling the temperature evolution (energy equation).

The concentration of the $i$-th chemical species can be also expressed in terms of the mass fraction $Y_i=\omega_i c_i / \bar \rho$, while, in an adiabatic closed system, the temperature is computed by conserving the mixture-averaged enthalpy, which for ideal gases reads 
\begin{equation}\label{mixture.averaged.enthalpy}
\bar h = \sum\limits_{i = 1}^n {Y_i h_i \left( \theta \right) },
\end{equation}
where $\bar \rho$, $\omega_i$ and $h_i$ are the mixture density, the molecular weight and specific enthalpy (per unit mass) of species $i$, respectively. 
For the sake of completeness, we report here the closed dynamical system governing closed reactive ideal mixtures under fixed enthalpy $\bar h$ and pressure $p$ to be addressed below in section \ref{illustration.h2o2}:
\begin{equation}\label{detailed.system.hp}
\left\{ \begin{array}{l}
 d\psi /dt = \sum\limits_{s = 1}^r {\gamma _s W_s \left( {\psi ,\theta } \right)}  = \left( d c_1/dt, ..., dc_n/dt  \right)\\ 
 d\theta /dt =  - \frac{1}{{\bar C_p }}\sum\limits_{i = 1}^n {h_i \left( \theta  \right)\dot Y_i }  \\ 
 \end{array} \right.
\end{equation}
where $\bar C_p$ denotes the mixture-averaged specific heat under fixed pressure, while specific enthalpy $h_i \left( \theta  \right)$ for any species $i$ can be computed using (\ref{polyprop}). Molar concentrations $c_i$ are linked to mass fractions $Y_i$ as $c_i=p  \left( Y_i / \omega_i  \right) /  \left( \mathcal{R} \theta \sum_j ^n Y_j / \omega_j  \right)$, while the mass fraction rate $\dot Y_i$ reads as follows: $\dot Y_i= \omega_i \bar \rho^{-1} d c_i/dt$, where $\omega_i$ is the molecular weight of species $i$. We notice that the second equation in (\ref{detailed.system.hp}) stipulates the conservation of $\bar h$, thus it represent an alternative way of imposing constance of (\ref{mixture.averaged.enthalpy}).

Finally, due to the conservation of elements, in a closed reactor, $d$ linear combinations of the species concentrations (expressing the number of moles of each element) remain constant during the system evolution in time:
\begin{equation}\label{element.laws}
C \psi=const,
\end{equation}
where $C$ is a $d \times n$ fixed matrix.

{\em Remark}--Having in mind dissipative multiscale dynamics such as chemical and physical kinetics, here we focus on systems (\ref{kin_sys}) with a single steady state. Hence, the Relaxation Redistribution Method (RRM) introduced below in section \ref{RRM.gen.sec} has been tested for those cases so far. We stress however that, for deriving the RRM approach, no assumptions are made concerning the number of steady state points of (\ref{kin_sys}). Thus, implementations of the RRM to different dynamics shall be presented in future publications.
\subsection{Thermodynamic Lyapunov function}\label{thermo.G.function}
Due to the second law of thermodynamics, the kinetic equations (\ref{mol.concentration.eq}) are equipped with a global thermodynamic Lyapunov function $G \left( \psi \right)$. 
In other words, the time derivative of the above state function is non-positive in the whole phase-space, $\dot G \left( \psi \right) \le 0$, with the equality holding at steady state.
For instance, in an adiabatic reactor with fixed pressure $p$ and enthalpy $\bar h$, the specific mixture-averaged entropy $\bar s$ (in mass units) monotonically increases in time starting from any non-equilibrium initial condition: hence the function $G = - \bar s$ decreases during the dynamics. For ideal gas mixtures, a Lyapunov function $G$ of the system (\ref{mol.concentration.eq}) takes the explicit form:
\begin{equation}\label{lyap.func}
G = -\bar s = - \sum_{i=1}^n X_i \left[ s_i \left( \theta \right) - \mathcal{R} {\rm ln} X_i - \mathcal{R} {\rm ln} \left(p/p_{ref}\right) \right] /\bar W,
\end{equation} 
where $X_i = c_i / \sum_{i=1}^n c_i$ and $s_i$ denote the mole fraction and the specific entropy of species $i$, respectively, $\mathcal{R}$ is the universal gas constant, $p_{ref}$ a reference pressure and $\bar W$ the mean molecular weight. For numerical purposes, the properties of the $i$-th species, $h_i$ and $s_i$, can be expressed in terms of the temperature, $\theta$, and a set of tabulated coefficients $a_{ij}$ as follows \cite{chemkin}:
\begin{equation}\label{polyprop}
\begin{array}{l}
h_i \left( \theta \right) = \mathcal{R} \theta \left( a_{i1} + \frac{a_{i2}}{2} \theta + \frac{a_{i3}}{3} \theta^2 + \frac{a_{i4}}{4} \theta^3 + \frac{a_{i5}}{5} \theta^4 + \frac{a_{i6}}{\theta} \right), \\
s_i \left( \theta \right) = \mathcal{R} \left( a_{1i} {\rm ln} \theta + a_{i2} \theta + \frac{a_{i3}}{2} \theta^2 + \frac{a_{i4}}{3} \theta^3 + \frac{a_{i5}}{4} \theta^4 + a_{i7} \right).
\end{array}
\end{equation}

\section{The film equation of dynamics}\label{film.equation.sec}
If the number of degrees of freedom $n$ is large, one may seek a reduced description with a smaller number of variables $q\ll n$. A consistent approach to model reduction is provided by the Method of Invariant Manifold (MIM) whose brief review is in order. Interested reader is delegated to the work \cite{book} for further details.

\begin{figure}[htbp!]
	\centering
		\includegraphics[width=0.85\textwidth]{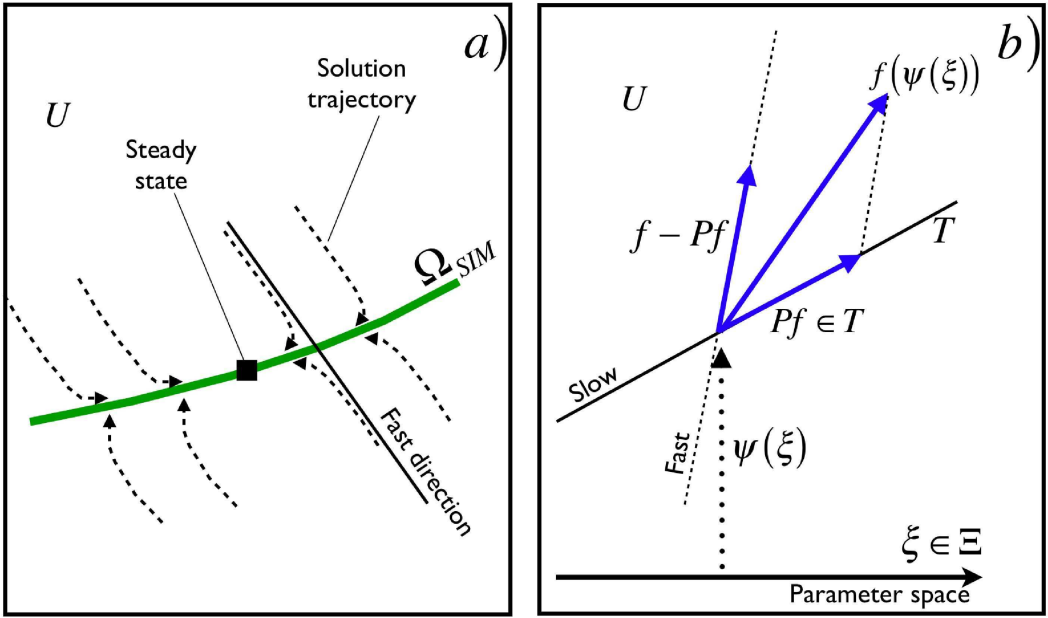}\\
	\caption{(Color online). a) Model reduction techniques assume the following idea: After a fast initial transient at the time instants $t \le t_0$, the (slow) dynamics of a complex system takes place along a slow invariant manifold (SIM) on the phase-space $U$ at any future time $t>t_0$ (invariance) toward the steady state. b) The definition of a projector $P$ onto the tangent space $T$ introduces a decomposition of slow and fast motions of the field $f$. In a vicinity of the SIM, slow and fast motions are locked in the image and null space of the thermodynamic projector $P$ \cite{book,CiCP2010}, respectively.}\label{cartoon_slowfast}
\end{figure}

In MIM, the problem of model reduction is identified with the construction of  a slow invariant manifold (SIM) $\Omega_{\rm SIM}$, whose dimension $q$ is the number of the essential (macroscopic) variables which parameterize the SIM. 
As sketched in the cartoon in Fig. \ref{cartoon_slowfast}a), the above method is based on the idea that the macroscopic {\em slow} dynamics of a complex system occurs along the SIM (invariance), once an initial {\em fast} relaxation toward the SIM has taken place. 
Let a manifold $\Omega$ (not necessarily a SIM)  be  embedded in the phase space $U$ and defined by a function $\Omega=\psi(\xi)$ which maps a macroscopic variables space $\Xi$ into $U$. Introducing a projector $P$ onto the tangent space $T$ of a manifold $\Omega$, the reduced dynamics on it is defined by the projection  $P f(\Omega) \in T$ (see Fig. \ref{cartoon_slowfast}b)). 
A manifold $\Omega$ is termed invariant (but not necessarily {\it slow}) if the vector field $f$ is tangent to the manifold at every point:
$f(\psi(\xi))- P f(\psi(\xi)) = 0, \; \xi \in \Xi$.

While the notion of a manifold's invariance is relatively straightforward, a definition of slowness is more delicate as it necessarily compares a (faster) approach towards the SIM with a (slower) motion along SIM. In MIM, slowness is understood as {\em stability}, and SIM is a stable stationary solution $\psi_{\rm SIM}(\xi)$ of the following {\em film equation} of dynamics defined on the space of maps $\psi(\xi)$ \cite{book},
\begin{equation}\label{film.dynamics}
\frac{d \psi({\xi})}{dt} = f (\psi(\xi))- P f (\psi(\xi)).
\end{equation}
Rigorous proofs of existence and uniqueness of SIM, by the film equation (\ref{film.dynamics}), were recently given for linear systems \cite{Ramm09}, while the rationale behind the (\ref{film.dynamics}) is explained by means of a cartoon in the Fig. \ref{cartoon01}a).
Here, it is worth stressing that the above (\ref{film.dynamics}) denotes a partial differential equation (PDE) whose unknown is a mapping $\psi \left( \xi \right)$ from a low dimensional reduced space $\Xi$ - $\xi \in \Xi$ - (also referred to as {\em parameter space} in the following) into the phase space $U$ - $\psi \in U$. Therefore, readers should not get confused between stable stationary solutions of (\ref{film.dynamics}) (defining SIM as a mapping from $\Xi$ into $U$) and single stationary states (or equilibrium states) of (\ref{kin_sys}) $\psi^{ss}$ (which satisfy the condition: $f \left( \psi^{ss} \right)=0$).

For thermodynamically consistent systems (\ref{kin_sys}) equipped with a potential $G$ (thermodynamic Lyapunov function with respect to (\ref{kin_sys})), MIM offers a projector whose construction is based on the tangent space $T$ and the gradient of the thermodynamic potential, $\partial G/ \partial \psi$, at every point of SIM. This consistently imposes that the reduced dynamics $P f (\psi(\xi))$ is dissipative. Explicit formulae for this thermodynamic projector are not necessary for the scope of this paper, and can be found in \cite{book}.
%
Importantly, separation of motions in a vicinity of SIM is dictated by thermodynamic projector $P$, since it can be proved that slow motions along SIM are locked in the image, ${\rm im}P=T$, whereas the null-space, ${\rm ker}P$, spans the fibers of fast motions transversal to SIM (Fig. \ref{cartoon_slowfast}b)) \cite{CiCP2010}.

Finally, a computationally advantageous realization is provided by a {\em grid} representation of MIM \cite{GK03}, where grid nodes in the phase space  are defined by a discrete set of macroscopic variables, $\xi$, while finite difference operators are used to compute the tangent space at every node $\psi (\xi)$.
Thanks to locality of MIM constructions, we further make no distinction between manifolds and grids.

{\em Remark}--Consistent constructive methods of slow invariant manifolds rely upon efficient methods for solving the PDE (\ref{film.dynamics}). As discussed below in section \ref{direct.filmeq}, towards this aim, finite difference schemes have been suggested in the literature \cite{GKZ04,book,Nafe02} (see also (\ref{relaxation.scheme})). Nevertheless, to the best of our knowledge, only explicit (or semi-implicit) schemes are available so far. Thus, due to hyperbolicity of the equation (\ref{film.dynamics}), its numerical solution is hindered by numerical instabilities (i.e. Courant type) \cite{GKZ04}, and no satisfactory solution to this issue has been suggested up to now. It is useful to stress that here we review the notion of film equation only for a better understanding of the present work. In fact, our suggestion toward the effective answer to the above problem is to avoid direct solution of (\ref{film.dynamics}) (e.g. by finite difference schemes) in favor of its {\em emulation}, where the problematic term $-Pf$ is not approximated with finite differences but mimicked by a {\em redistribution} step in terms of macroscopic variables (see section \ref{RRM.gen.sec} below).

\subsection{Direct solution of the film equation}\label{direct.filmeq}
A natural approach to the construction of SIM's is a direct numerical solution of the film equation (\ref{film.dynamics}) starting with an initial (usually non invariant) manifold. For that, both the initial condition as well as implicit or semi-implicit schemes were developed. The simplest explicit scheme for solving the equation (\ref{film.dynamics}) can be realized by iteratively refining each point $\psi$ of the initial manifold: $\psi + d\psi$,
\begin{equation}\label{relaxation.scheme}
d\psi  = \tau \left( {f\left( \psi  \right) - Pf\left( \psi  \right)} \right),
\end{equation} 
with the time $\tau$ being estimated according to the suggestions in \cite{GK03}, where the scheme (\ref{relaxation.scheme}) is referred to as the {\em relaxation method}. It has been noticed \cite{GKZ04} that the solution of the film equation of dynamics (\ref{film.dynamics}), similarly to hyperbolic partial differential equations for computational fluid dynamics (CFD) simulations, is hindered by severe numerical instabilities (see, e.g., the Courant instability \cite{Cou67}). Furthermore, we notice that, unlike CFD, numerical solution of (\ref{film.dynamics}) comes with additional difficulties, due to an uncontrolled variation of the grid-node spacing. As a result, it is difficult to formulate an analog of the CFL (Courant - Friedrichs - Lewy) condition \cite{Cou67} for (\ref{relaxation.scheme}), and the suppression of instability was only attempted by an arbitrary decrease of the time $\tau$ until convergence \cite{GKZ04}. In general, the latter approach proves rather poor since the lack of convergence of (\ref{relaxation.scheme}) might not have numerical origin. In fact, there is no guarantee that the chosen number of reduced degrees of freedom $q$ reveals sufficient in describing the asymptotic behavior of the dynamical system (\ref{kin_sys}) in a given domain of the phase-space. For instance, in the case a higher number of reduced variables are requested, the refinement of a $q$ dimensional manifold by {\em stable} numerical schemes of (\ref{film.dynamics}) is expected to fail anyway. The idea of adaptive dimension of SIM, formulated below in the section \ref{RRM.local.sec}, is based on the latter observation.

Finally, the construction of slow invariant manifolds by the solution of (\ref{film.dynamics}) has been always attempted in the whole phase-space, by assigning {\em a priori} their dimension $q$ somewhat arbitrarily. Such an approach, where the dimension $q$ comes as {\em external input} into the problem, poses severe limitations to the accuracy of the reduced description and, most detrimentally, hinders the gaining of any better physical knowledge about it. Moreover, construction of high-dimensional invariant manifolds ($q \ge 3$) by the (\ref{film.dynamics}) is quite problematic and was never successfully accomplished up to now.


\section{The relaxation redistribution method: RRM}\label{RRM.gen.sec}

Toward the end of overcoming the above drawbacks, in this work, we introduce an approach to model reduction, which allows for the construction of slow invariant manifolds with the dimension $q$ adaptively varying from one region of the phase space to another. We address thereby the fundamental issue of the {\em minimal} number of important (slow) variables which underlie the behavior of a complex dissipative phenomenon in a region of the phase space: A knowledge, emerging from the system and no longer imposed, is now gained. The latter is a challenging problem in Physics, and even in the classical cases, such as the reduced description of the Boltzmann kinetic equation by a finite set of velocity moments of the distribution function (see, e.g., \cite{MullerRuggeribook}), some essential questions remain open \cite{KCK08,OttingerPRL,Struchtrupbook}. Similarly, in chemical kinetics, several methods have been suggested \cite{KeG71,Lebiedz04,RenPope06} for approximating and parameterizing the SIM, however the choice of the minimal number of chemical coordinates (manifold parameters) is still debated.

In the following, the key idea of our approach is to abandon an attempt of {\it solving} the film equation (\ref{film.dynamics}) by numerical schemes such as (\ref{relaxation.scheme}), in favor of a {\it simulation} of the physics behind this equation, in a spirit similar to Monte Carlo methods: As a consequence, a highly efficient construction of SIM with an embarrassingly simple implementation in any dimension is derived.
\subsection{Global formulation of RRM}\label{RRM.global.sec}
\begin{figure}[htbp!]
	\centering
		\includegraphics[width=0.85\textwidth]{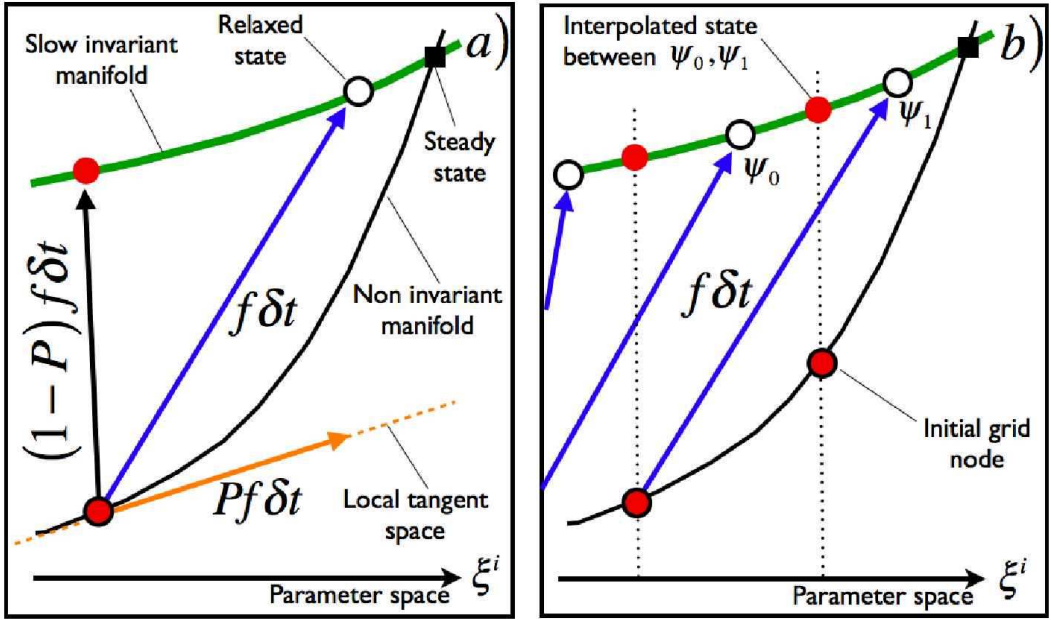}\\
	\caption{(Color online). a) The relaxation due to (\ref{kin_sys}) of a non-invariant manifold. Fast dynamics drives it toward the slow invariant manifold, whereas the concurrent action of the slow dynamics causes a shift toward the steady state (shagreen effect). On the contrary, relaxation due to the film equation (\ref{film.dynamics}) - (\ref{relaxation.scheme}) allows movements only in the fast subspace. b) Relaxation Redistribution Method. The displacement in the slow subspace, generated during relaxation, is annihilated by a redistribution step in the parameter space.}\label{cartoon01}
\end{figure}
In order to introduce our method, we consider reaction kinetics and assume that
a  slow dynamics of (\ref{kin_sys}) evolves on a $q$-dimensional SIM in the $n$-dimensional concentration space
(this assumption will be relaxed in a sequel).
Inspection of the right-hand side of (\ref{film.dynamics}) reveals a composition of two motions: The first term, $f(\psi(\xi))$, is the relaxation of the initial approximation to SIM due to the detailed kinetics, while the second term, $-P f( \psi (\xi))$ is the motion {\it antiparallel} to the slow dynamics.
Let a time stepping $\delta t$ and a numerical scheme (e.g. Euler, Runge-Kutta, etc.) be chosen for solving the system of kinetic equations: All grid nodes relax towards the SIM under the full dynamics $f$ during $\delta t$.
Fast component of $f$ leads any grid node closer to the SIM while at the same time, the slow component causes a shift towards the steady state (see Fig. \ref{cartoon01}a)).
As a result, while keeping on relaxing, the grid shrinks towards the steady state (we term this a "shagreen effect"  per {\it de Balzac}'s famous novel \cite{Balzac} - chagrin in French). Subtraction of the slow component therefore prevents the shagreen effect to occur, and it is precisely the difficulty in the numerical realization: explicit evaluation of the projector $P$ on the approximate SIM does not always balance the effect of shrinking. This leads to instabilities, and results in a drastic decreasing of the time step.

The key idea here is to neutralize the slow component of motion by a {\em redistribution} of the points on the manifold  after the relaxation step (see Fig. \ref{cartoon01}b)).
For the sake of presentation, we assume that macroscopic parameters are given by a set of $q$ linear functions $b=\{b_1,\dots,b_q\}$ such that $b_1(\psi)=\xi^1$, $\dots$, $b_q(\psi)=\xi^q$.
Let $\xi=\left( {\xi^1 , \ldots ,\xi^q } \right)$ be a generic node of a fixed grid $S$ in the parameter space $\Xi$, and the $q$-dimensional slow invariant grid (SIG) in the phase space $U$ is initialized: 
 $\Omega^{\rm in}=\psi^{\rm in}(\xi)$ 
 (that is, the initial SIG is the collection of nodes $\psi^{\rm in}=\psi^{\rm in}(\xi)$, $\xi\in S$). 
After the relaxation step, all the nodes $\psi^{\rm in}$ have moved to new locations, $\psi^{\rm in}\to\psi^{\rm R}$, and we denote $\xi^{\rm R}=b(\psi^{\rm R})$ the values of the macroscopic parameters corresponding to the relaxed nodes $\psi^{\rm R}$. 

It is worth stressing that by {\em parameter space} here we mean the low dimensional macroscopic space $\Xi$ whose dimension is $q<<n$. Hence, an arbitrary grid $S$ is defined by a mapping, $\psi \left( \xi \right)$, on a subspace of $\Xi$ into the phase-space $U$ (of dimension $n$).

For example, the forward Euler scheme used below gives
\begin{equation}
  \psi^{\rm R}=\psi^{\rm in}(\xi)+\delta t f(\psi^{\rm in}(\xi)).
 \end{equation}
With this, also the nodes of the grid $S$ shift by an amount $\delta\xi=b(\psi^{\rm R})-\xi$ due to the slow component of motion.
The redistribution of the nodes $\psi^{\rm R}$ back to the fixed grid $S$ simulates the subtraction of the slow motion from the relaxation step, and is done as follows: 
For each $\xi\in S$, we consider a $q$-simplex $S_q$ (in $U$) with $q+1$ vertices 
$\psi^{\rm R}_{0}$, $\psi^{\rm R}_{1}$, $\dots$, $\psi^{\rm R}_{q}$ such that $\xi$ is inside the macroscopic projection of $S_q$, the simplex $\Sigma_q$ (in $\Xi$) formed by the vertices 
$\xi^{\rm R}_{0}=b(\psi^{\rm R}_{0})$, $\xi^{\rm R}_{1}=b(\psi^{\rm R}_{1}$), $\dots$, $\xi^{\rm R}_{q}=b(\psi^{\rm R}_{q})$.
The updated (relaxed-and-redistributed) grid $\Omega^{\rm RR}$ is constructed by a linear interpolation of the vertices of the simplex $S_q$: 
%
%
%
\begin{equation}\label{lin.interp}
\psi^{\rm RR} = \left(1 - \sum_{i = 1}^q w_{i} \right)\psi^{\rm R}_{0}  + \sum_{i = 1}^q {w_{i} \psi^{\rm R}_{i} } ,
\end{equation}
where the weights $w_i$ are so chosen as to satisfy the redistribution condition, 
\begin{equation}\label{RRcondition}
b(\psi^{\rm RR})=\xi.
\end{equation}
This amounts to solving a $q\times q$ linear system,
\begin{equation}\label{Lagrange}
\sum_{j=1}^{q}[b_j(\psi^{\rm R}_i)-b_j(\psi^{\rm R}_0)]w_j=\xi^i-b_i(\psi_0^{\rm R}).
\end{equation}
The above procedure is supplemented by the boundary conditions applied at the edges of the grid: Grid nodes at the boundary $\psi_{\rm b}$ are reconstructed by extrapolation after the relaxation step. Formula (\ref{lin.interp}) is used where $\psi^{\rm RR}=\psi_{\rm b} \notin S_q$ is located in the vicinity of a simplex $S_q$ with vertices $\psi^{\rm R}_0,\psi^{\rm R}_1, \dots,\psi^{\rm R}_q$.
In general, $S_q$ can be chosen in such a way that its vertices are the relaxed states of the initial nodes $\psi^{\rm in}_0,\psi^{\rm in}_1, \dots,\psi^{\rm in}_q$ with $\psi_{\rm b}=\psi^{\rm in}_0$.
%

Thus, after the redistribution step, the initial grid is refined towards the invariant grid. The procedure is then iterated, whereas each relaxation step is altered by the redistribution step, in which the slow motion is subtracted by stretching the macroscopic variables to the nodes of the initial grid $S$.

We notice that, on SIM, movements due to the vector field $f$ occur along the manifold itself, thus the effect of the relaxation is entirely counterbalanced by the subsequent redistribution on the SIM. It is worth stressing that this observation holds for every invariant manifold (not necessarily SIM). Nevertheless, numerical evidences clearly show that an arbitrary invariant manifold $\Omega_{inv}$ is an {\em unstable} solution of the above dynamics, and refinements starting from $\Omega_{inv}$ converge toward the SIM, which instead turns out to be a {\em stable} solution. 
As a result, slow invariant grids are {\em stable stationary solutions} of the described procedure, here termed \textit{relaxation redistribution method} (RRM).
Once the invariant grid is constructed, the reduced dynamics for variables $\xi$ is defined as 
\begin{equation}\label{macro}
\frac{d\xi}{dt}=b(f(\psi^{\rm RR}(\xi)).
\end{equation}
In other words, the suggested RRM enables to provide the reduced system (\ref{macro}), written in terms of a significantly smaller set of variables $\xi$, with a {\em closure}.  

Note that, upon the global construction of SIG, computations deliver a discrete set of {\em linked} states $\psi^{\rm RR}(\xi)$, in a vicinity of the corresponding slow invariant manifold. Here, grid nodes are termed "interconnected" because we assume that for any arbitrary node it is possible to identify all its nearest neighbors. Moreover, interconnectivity enables one to easily proceed with analytical continuation of the above slow invariant grid, and thus to the calculation of the right-hand side of (\ref{macro}) for any set of variables $\xi$. To this end, for simplicity, here we adopt multi-linear interpolation, which posses the advantage to automatically fulfill the linear conservation constraints (\ref{element.laws}). For further details on multi-linear interpolation of grids, the interested reader is delegated to \cite{ChiavazzoPHD}. On the other hand, if the local construction of SIG is implemented, a closure for (\ref{macro}) is computed when needed and no analytical continuation of the grid is requested. In the latter case, in order to speed up the computations, smart methodologies for data storage and retrieval can be used and are readily available from the literature (see, e.g., the ISAT method in \cite{Pope97}).

Finally, note that while the redistribution step seems "natural" from the numerical standpoint of discretizing the above macroscopic equation (\ref{macro}) on a fixed grid $S$, the feature recognized here is that it is precisely the subtraction of the slow component of the motion in the film equation (\ref{film.dynamics}), which circumvents the question about explicit evaluation of slow motions in the course of the SIM construction.

\begin{figure}[htbp!]
	\centering
		\includegraphics[width=0.85\textwidth]{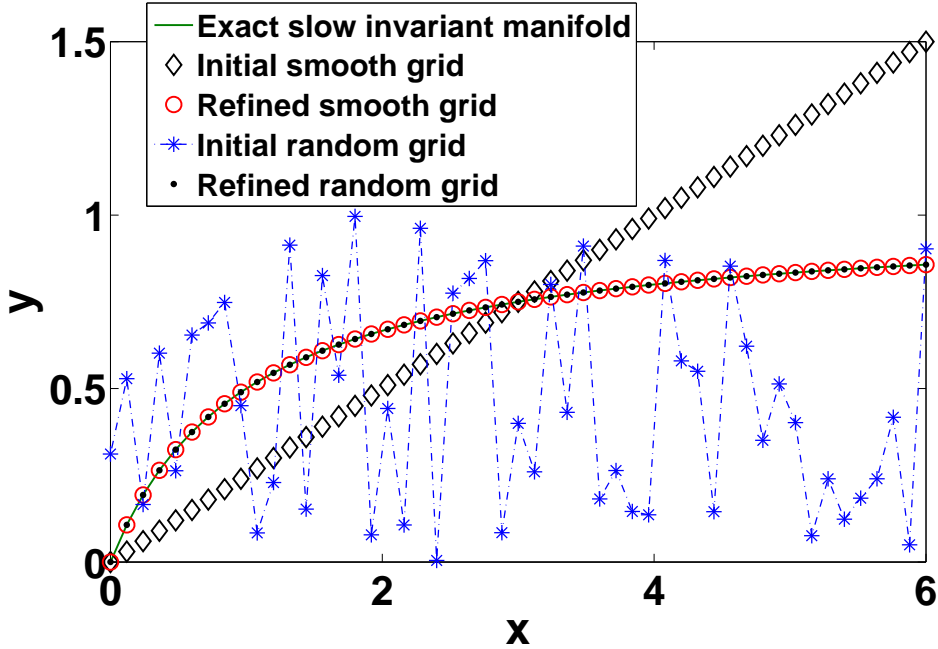}\\
	\caption{(Color online). The Davis-Skodje system \cite{DS99}.
%
Two different initial grids are refined using the forward Euler scheme for the relaxation ($\delta t=10^{-2}$). Results after $50$ RRM iterations are reported (refined grids) with $\gamma=50$. Triangles show an intermediate step (after two RRM iterations) starting from the initial smooth grid.}\label{Fig1}
\end{figure}
In order to test the RRM, we first consider a simple benchmark suggested by Davis and Skodje (DS) \cite{DS99} (a two-dimensional system with a one-dimensional SIM known in a closed analytical form). 
The DS system \cite{DS99} consists of two equations,
\begin{equation}
\begin{array}{l}
 {{dx} \mathord{\left/
 {\vphantom {{dx} {dt}}} \right.
 \kern-\nulldelimiterspace} {dt}} = f_x \left( x \right) =  - x, \\ 
 {{dy} \mathord{\left/
 {\vphantom {{dy} {dt}}} \right.
 \kern-\nulldelimiterspace} {dt}} = f_y \left( {x,y} \right) =  - \gamma y + {{\left[ {\left( {\gamma  - 1} \right)x + \gamma x^2 } \right]} \mathord{\left/
 {\vphantom {{\left[ {\left( {\gamma  - 1} \right)x + \gamma x^2 } \right]} {\left( {1 + x} \right)^2 ,}}} \right.
 \kern-\nulldelimiterspace} {\left( {1 + x} \right)^2 ,}}\quad \gamma  > 0 \\ 
 \end{array}
\end{equation}
has unique stable steady state $x=y=0$, and a one-dimensional SIM, $y=x/(1+x)$. Here, when $\gamma >> 1$, due to a significant separation between of time scales of the two variables $x$ and $y$, all solution trajectories of DS system exponentially decay to the SIM (see Ref.\ \cite{DS99}). 
In the above notation, $\psi=(x,y)^T$, and we define the slow variable as $\xi=x$, that is, $b=(1,0)$ and $b(\psi)=(1,0)\cdot(x,y)^T=x$. The RRM is initialized with the grid represented by the collection of points $\{(x_{r},y^{\rm in}(x_r))\}$, where $x_r$ are distributed evenly in the interval $x\in [0,x_{\rm b}]$. Upon the relaxation step, the grid points are shifted to new locations
$\{(x_{r},y^{\rm in}(x_r))\}\to\{(x^{\rm R}_{r},y^{\rm R}(x_r))\}$ with $x_r^{\rm R}=(1-\delta t)x_r$, $y_r^{\rm R}=y^{\rm in}(x_r)+\delta t f_y(x_r,y^{\rm in}(x_r))$. Choosing the interval $\Sigma_1=[x_0^{\rm R},x_1^{\rm R}]$ for each point $x_r$ such that $x_r\in \Sigma_1$, the redistribution (\ref{lin.interp}) gives
\begin{equation}
y^{\rm RR}_r=\frac{(x_1^{\rm R}-x_r)y_0^{\rm R}+(x_r-x_0^{\rm R})y_1^{\rm R}}{x_1^{\rm R}-x_0^{\rm R}},
\end{equation}
while $x_r^{\rm RR}=x_r$, by the condition (\ref{RRcondition}).
For the boundary node at $x_{\rm b}$ we set $y_0^{\rm R}=y_{\rm b}^{\rm R}$ and for $y_1^{\rm R}=y^R(x_{b-1})$, with $x_{b-1} \in S$ being the nearest neighbor of $x_b$. In Fig. \ref{Fig1}, the local grid step at the boundary is: $\delta x_b=x_b-x_{b-1}=6-5.88$.

The RRM was performed for a variety of initial grids (initialized with different functions $y^{\rm in}(x)$), with different spacing and with various choices of the simplex. Independent of these variations, the RRM iterations converged stably to the analytical SIM of the DS system.
Results are presented in Fig.\ \ref{Fig1} for two different initial grids, a regular linear ($y^{\rm in}=ax$, $a=0.25$) and a randomly generated grid (for each value $x_r \in [0,6]$ a random number $y^{\rm in}(x_r)  \in  ]0,1[ $ is assigned by a linear congruent generator) with the intervals $\Sigma_1$ chosen as $x_1-x_0=0.12$. Convergence to SIM is even striking in Fig.\ \ref{Fig1} given the fact that both initial grids are far from SIM.

Thus, convergence of the RRM iterations confirms the existence of a reduced description with a fixed number of degrees of freedom $q$ (existence of $q$-dimensional slow invariant manifold). On the contrary, no convergence in RRM indicates that more degrees of freedom are needed to recover the detailed system dynamics. 
This concept shall be used below for adaptively choosing the invariant grid dimension.

Both construction and usage of a global reduced description soon become impracticable as the dimension $q$ increases. In fact, computing and storage of high dimensional SIM's may be problematic already at $q \ge 3$. Above all that, data retrieval by interpolation on such large arrays is computationally intensive, and sometimes full construction of manifolds can be useless: For example, in combustion applications, regions with high a concentration of radicals are unlikely to be visited.

%

{\em Remark}--In general, when using model reduction techniques, such as the RRM method, slow and fast subspaces are not known in advance. 
In fact, this kind of information is what we get at the end of the process. Invariant grids 
constructed by the suggested RRM are 
finally located in the slow subspace (regardless of the choice on the 
parameterization). The fast subspace can be thereafter reconstructed by adopting 
e.g. the notion of thermodynamic projector (see, e.g., Refs. \cite{book,GKZ04,CiCP2010}). On the other side, 
concerning the parameterization choice, we notice that (as stressed in the 
conclusions \ref{conclusion.sec}) there are no universal recipes, and it specifically 
depends on the physical phenomenon we are dealing with. In general Ògood 
macroscopic variablesÓ can be found in the literature: For instance, in the case of 
the Boltzmann equation typical macroscopic parameters are the velocity 
moments of the distribution function, whereas for chemical kinetics we can use 
spectral variables as done for the example in section V. Alternatively, in the latter 
case, typical slow variables can also be adopted (see, e.g., the RCCE 
parameterization in \cite{KeG71}).

\subsection{Local formulation of RRM}\label{RRM.local.sec}
\begin{figure}[htbp!]
	\centering
		\includegraphics[width=0.85\textwidth]{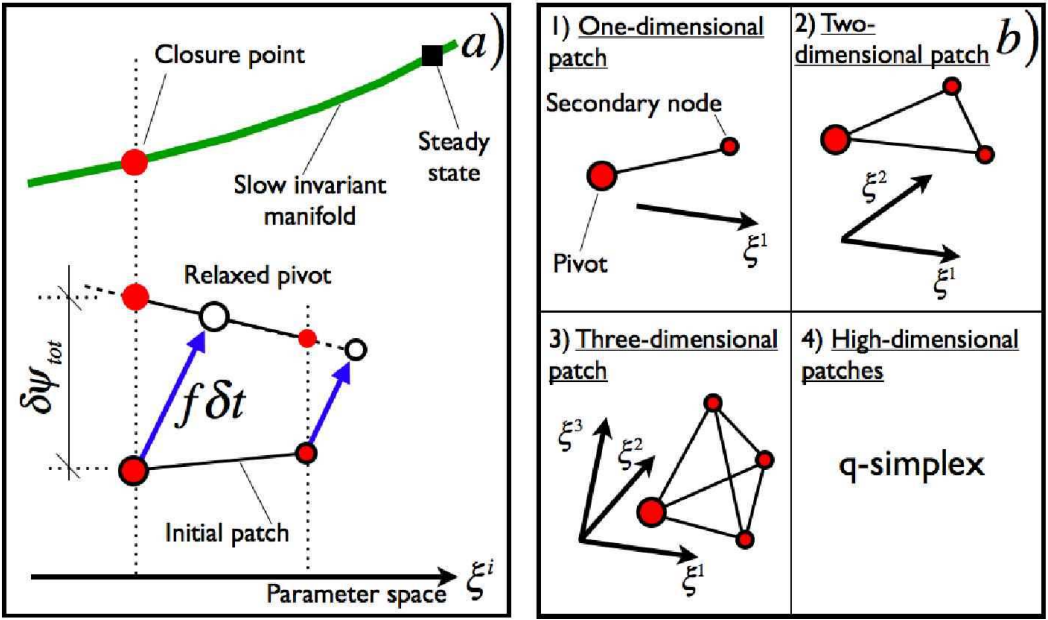}\\
	\caption{(Color online). a) Relaxation redistribution method: Local formulation. Only a small patch of the SIM is constructed. After refinement, the coordinates of the pivot provide the reduced system (\ref{macro}) with a closure. b) Simplexes can be conveniently adopted for a patch-wise description of the SIM in any dimension.}\label{cartoon02}
\end{figure}
Importantly, the RRM allows for a straightforward local formulation, where only small patches of the slow invariant grid are initialized and refined.
Let $\bar \xi=\left( {\bar \xi ^1 , \ldots ,\bar \xi ^q } \right)$ and 
the procedure is initialized with a simplex $\bar{S}_q$ where 
the pivot $\bar{\psi}^{\rm in}=\psi^{\rm in}(\bar{\xi})$
is linked to $q$ {\it secondary nodes} $\psi_1^{\rm in}={\psi}^{\rm in}( \bar \xi_1 ), ..., \psi_q^{\rm in}=\psi( \bar \xi_{q})$ in a neighborhood of $\bar{\psi}$ such that $\bar \xi _i  = \left( {\bar \xi ^1 , \ldots ,\bar \xi ^i  + \delta  \xi ^i , \ldots ,\bar \xi ^q } \right)$, with $\delta \xi^i$ being a small deviation of the $i$-th macroscopic variable. A sequence of relaxation and redistribution steps is applied to the vertices of $\bar{S}_q$ in any dimension $q$: This realizes indeed the simplest instance of the RRM,
 \begin{equation}\label{RRpivot}
\bar{\psi}^{\rm RR} = \left(1 - \sum_{i = 1}^q w_{i} \right)\bar{\psi}^{\rm R}  + \sum_{i = 1}^q {w_{i} \psi^{\rm R}_{i} },
\end{equation}
while the weights $w_i$ are found from the redistribution (anti-shagreen) condition (\ref{RRcondition}): $b(\bar{\psi}^{\rm RR})=\bar{\xi}$.
Refinements end as soon as a norm of the total displacement of the pivot at the $n$th RRM iteration, $\left|\delta\bar{\psi}^{\rm tot}_{(n)}\right|=\left|\bar{\psi}^{\rm RR}_{(n+1)}-\bar{\psi }^{\rm RR}_{(n)}\right|$, becomes sufficiently small compared to the displacement caused by the  relaxation alone, $\left| \delta\bar{\psi}^{\rm rel}_{(n)} \right|=\left|\bar{\psi}^{\rm R}_{(n+1)}-\bar{\psi }^{\rm RR}_{(n)}\right|$.

Setting an upper limit to both the number of refinements $N$ and the tolerance $\epsilon$ such that:
\begin{equation}\label{threshold}
\left| \delta\bar{\psi}^{\rm tot}_{(n)} \right| / \left| \delta\bar{\psi}^{\rm rel}_{(n)}\right|  \le \epsilon,
\end{equation}
the local RRM can be adaptively performed starting with $q=1$. If the latter requirements are not fulfilled, the dimension is updated to $q=2$ and the procedure repeated. Upon convergence with some $q=\bar q$, a closure of the reduced system (\ref{macro}) is provided by the coordinates of the pivot. 
It is worth stressing that the above convergence criterion (\ref{threshold}) is based on the value assigned to the tolerance $\epsilon$ and number of refinements $N$. However, the latter quantities can be properly set upon an independence study with respect to the manifold dimension. Namely, in the same spirit of grid independence studies of fluid dynamics simulation results, the independence of the manifold dimension $q$ on $\epsilon$ and $N$ can be verified by repeating the calculations with smaller tolerances and larger number of refinements.
In this sense, the local RRM fully alleviates any assumption about the dimensionality of SIM, the local dimension is found automatically and if no reduced description is possible at all, no convergence at any $q<n$ will clearly indicate this.

Finally, for systems supported by a Lyapunov functions $G$ (such as the kinetic equations (\ref{mol.concentration.eq})), a convenient (but not the only possible) initialization of the RRM procedure (construction of the initial pivot and secondary nodes) for dissipative systems can be accomplished by means of the notion of {\em quasi equilibrium manifold} (QEM). In this respect, an approximation of the $q$-dimensional SIM can  be obtained by minimizing the function $G$ under $q$ linear constraints in addition to the element conservation laws (\ref{element.laws}):
\begin{equation}\label{QEM.definition}
\left\{ \begin{array}{l}
 G \to \min  \\ 
 b_i \left( \psi  \right) = \xi ^ i ,\quad i = 1, \ldots ,q \\ 
 C\psi  = {\rm const}. \\ 
 \end{array} \right.
\end{equation}
where, in the case of chemical kinetics, the function $G$ is a thermodynamic potential (i.e., entropy, Gibbs free energy, etc.) as discussed in the section \ref{thermo.G.function}. It is worth stressing that the idea of using extrema of potentials, for providing reduced description with a closure, dates back to the work of Gibbs \cite{Gibbs}. From then on, this notion has been adopted is several areas such as the kinetic theory of gases \cite{Koganbook,book}, or detailed combustion mechanisms \cite{KeG71}. However, we stress that the latter approximations often provide with a poor description of the corresponding SIM \cite{ChK2007,RenPope06}, thus they are used here only for initializing the RRM.

Below, following the suggestion in \cite{ChGoKa07}, we make use of spectral variables $\xi^i=b_i(\psi)$ obtained by the inner product between the state $\psi$ and the parameterization vectors $b_i$, which are the left eigenvectors of the Jacobian $J=\partial f/\partial \psi$ at the steady state, corresponding to non-zero eigenvalues $\lambda_i$ and numbered in the order of increase of $|\lambda_i|$. The latter is referred to as {\em spectral quasi equilibrium} parameterization. The pivot  $\psi^* = \left( \psi_1,...,\psi_n \right)$ of the initial simplex  $\bar S_q$ is defined as the quasi-equilibrium point \cite{book}, corresponding to $\bar \xi = \left( {\bar \xi^1 , \ldots ,\bar \xi^q } \right)$, and calculated by solving the problem (\ref{QEM.definition}). To this end, the (\ref{QEM.definition}) is equivalent to the global minimization problem of a Lagrange function $\bar G$:
\begin{equation}\label{lagrange.function}
\bar G = G + \sum_{i=1}^q \left[ b_i \left( \psi  \right) - \xi^i \right] \tilde{\lambda}_i  + \tilde{\lambda} C \psi,
\end{equation}
with $\tilde{\lambda}_i$, $\tilde{\lambda}$ being a set of Lagrange multipliers. We notice that, efficient tools for the solution of (\ref{QEM.definition}) are also available (see, e.g., STANJAN \cite{Reyn86}).
Secondary nodes  $\psi^k$ of the simplex can be conveniently calculated by linear expansion of the minimization problem about the quasi-equilibrium as suggested in \cite{ChK2007}:  $\psi ^k  = \psi^* + \sum_{i = 1}^{n-d} {\delta _k^i \rho _i }$, with $\left( {\rho _1 , \ldots \rho_{n-d}} \right)$ and $\delta _k  = \left( {\delta _k^1 ,...,\delta _k^{n-d} } \right)$ being a vector basis spanning the null space of $C$ and the solution of a linear algebraic system
\begin{equation}\label{QEG.sys}
\left\{ \begin{array}{l}
 \sum\nolimits_{i = 1}^{n-d} {\left( {t_j H^* \rho _i} \right)\delta _i  =  - \nabla G^* t_j ,\quad j = 1, \ldots ,n-d-q},  \\
 \sum\nolimits_{i = 1}^{n-d} {\left( {b_1 \rho _i} \right)\delta _i  = 0},  \\
 \cdots  \\
 \sum\nolimits_{i = 1}^{n-d} {\left( {b_k \rho _i} \right)\delta _i  = \varepsilon _k },  \\
 \cdots  \\
 \sum\nolimits_{i = 1}^{n-d} {\left( {b_q \rho _i } \right)\delta _i  = 0}.  \\
 \end{array} \right.
\end{equation}
The vector basis $\left( {t_1 , \ldots t_{n - q - d} } \right)$ spans the kernel of the linear space defined by the vectors $b_i$ and the rows of the matrix $C$ in (\ref{element.laws}), $H^*=\left[ \partial^2 G /\partial \psi_i \partial \psi_j \right]$ and $\nabla G^* = \left[ \partial G / \partial \psi_i \right]$ are the second derivative matrix and the gradient of the function $G$ at the pivot respectively, while $\varepsilon_k$ defines the length of the edge of the simplex $\Sigma_q$ along the $k$-th direction.

\begin{figure}[htbp!]
	\centering
		\includegraphics[width=0.9\textwidth]{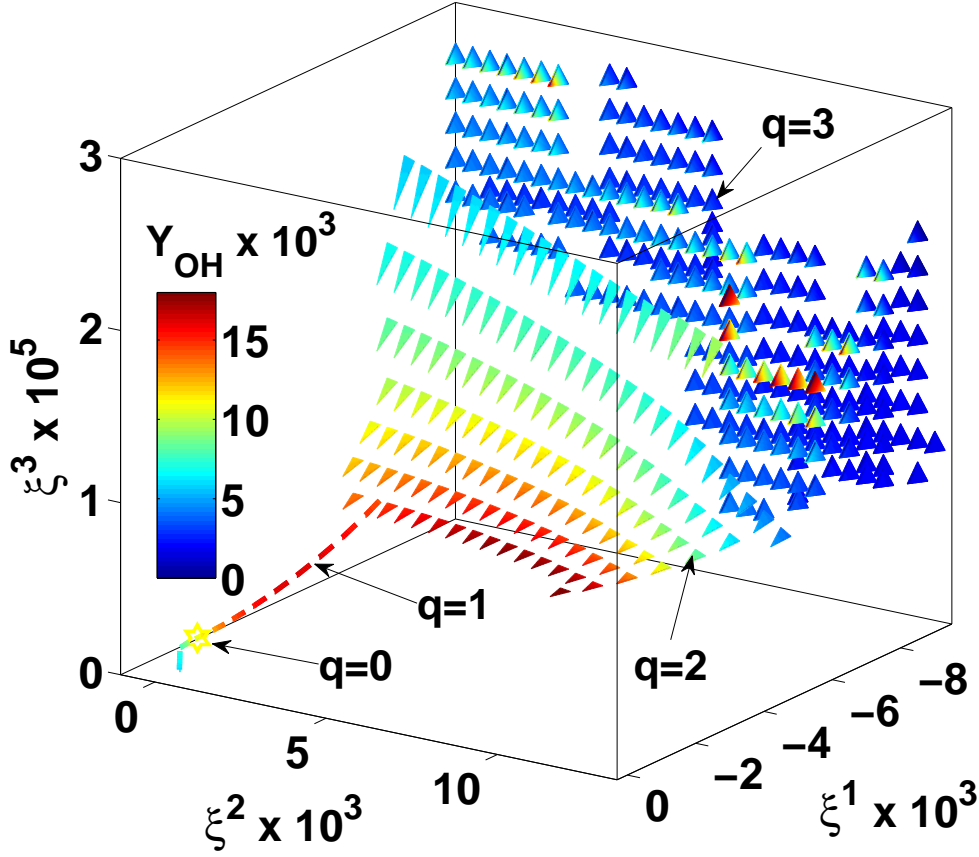}
	\caption{(Color online). Heterogeneous slow invariant manifold of hydrogen-air combustion mechanism by local RRM. Three-dimensional projection of the six-dimensional phase space onto spectral variables (see text). The two-dimensional patch ("kite", triangles) is tight by a one dimensional "thread" (line) to the zero-dimensional equilibrium and merges with the three-dimensional "cloud" (tetrahedra). Legend: mass fraction of OH. Explicit Euler scheme with $\delta t=5  \times 10^{-8} [s]$ was used for the relaxation of simplex. RRM convergence criteria: $N=2000$, $\epsilon=10^{-4}$.}\label{Fig2}
\end{figure}

\section{Illustration: Detailed hydrogen-air mixture}\label{illustration.h2o2}
Here, we considered the autoignition of hydrogen-air mixture at stoichiometric proportion, reacting according to the realistic detailed mechanism of Li et al. \cite{Limech}, where nine chemical species and three elements participate in a complex reaction dictated by twenty-one reversible elementary steps (\ref{elementary.step}) (this mechanism is universally used in turbulent combustion simulations \cite{NZZ}, and details for this case are discussed in the Appendix \ref{append01}). Time evolution of species concentration is governed by (\ref{mol.concentration.eq}), and it is supplemented by the condition for the reactor temperature, which stipulates the conservation of the mixture enthalpy (adiabatic reactor):
\begin{equation}
\bar h = \sum_{i=1}^9 h_i Y_i = 1000[kJ/kg]. 
\end{equation}
Furthermore, the pressure of the mixture is fixed ($p=1[atm]$), and the mass fraction ($Y_i$) of an arbitrary chemical species $i$ can be expressed in terms of the corresponding molar concentration ($c_i$) by means of the following relationship:
\begin{equation}\label{ideal.gas.law}
Y_i=\frac{c_i \omega_i}{\sum_{j=1}^n c_j \omega_j}.
\end{equation}

Fig.\ \ref{Fig2} shows a projection of the heterogeneous SIM (i.e. with a varying dimension in the phase-space), onto the subspace $\xi^1,\xi^2,\xi^3,Y_{OH}$, constructed by the local RRM where one-, two- and three-dimensional patches are clearly visible.
Here, the variables $\xi^1,\xi^2,\xi^3$ are chosen according to the spectral quasi equilibrium parameterization, where $\xi^i=b_i(\psi)=b_i \psi$ with $b_i$ denoting the three slowest eigenvectors of the Jacobian matrix $J = \partial f/\partial \psi$ at the steady state, whereas the RRM is initialized as discussed above in the text with the potential $G$ computed on the basis of the mixture-averaged entropy (\ref{lyap.func}). Interested reader may find full details on the computation of $G$ and its derivatives ($\nabla G$ and $H^*$ requested in (\ref{QEG.sys})) in \cite{CF2010}. Results in terms of basic variables (i.e. concentrations of species) can be obtained upon a post-processing of the spectral variables which amounts to a linear transformation.

\begin{figure}[htbp!]
	\centering
		\includegraphics[width=0.9\textwidth]{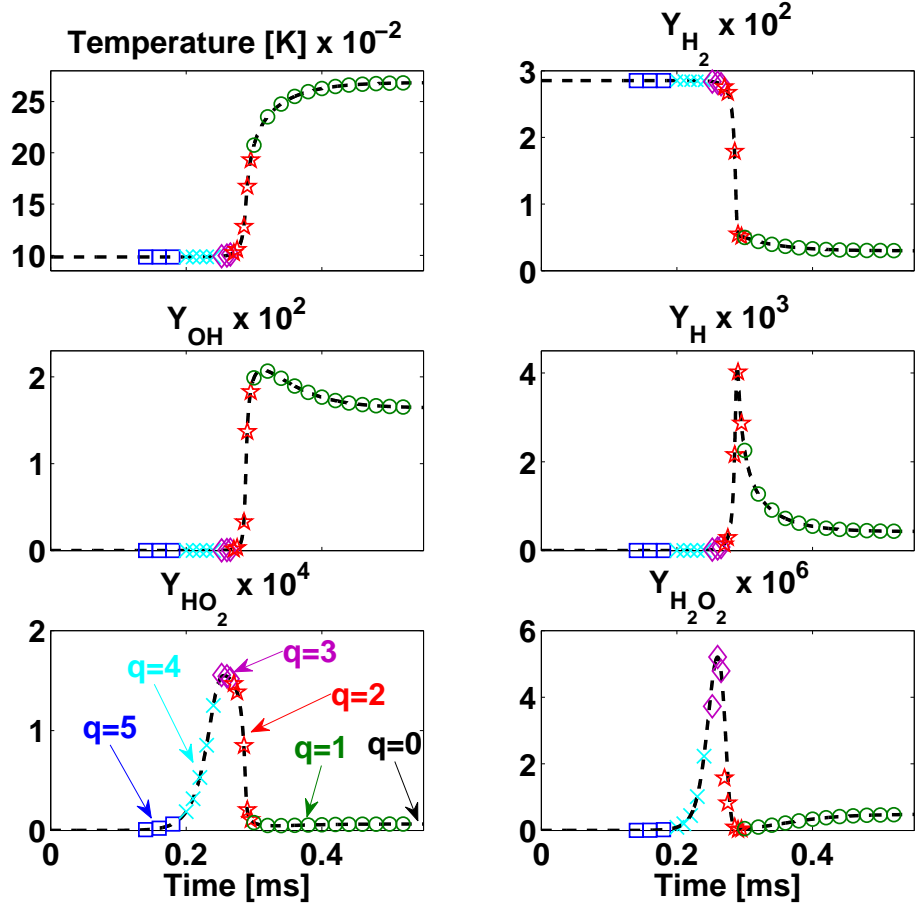} 
	\caption{(Color online). Auto-ignition of homogeneous stoichiometric mixtures of hydrogen and air: Histories of the temperature and of the mass fraction of chemical species. Line: detailed reaction; Symbol: local RRM method by adaptively following a cascade of reduced models of various dimension $q$: $q=5$ (square), $q=4$ (cross), $q=3$ (diamond), $q=2$ (star), $q=1$ (circle) and $q=0$ (steady state).}\label{Fig3}
\end{figure}

A typical problem, where dynamics evolves along a {\em cascade} of slow invariant manifolds with progressively lower dimensions, is the auto-ignition of a fuel-air mixture.
In Fig. \ref{Fig3}, solution of the reduced system (\ref{macro}), supplemented with a closure by the local formulation of RRM, is compared with the integration of the detailed reaction mechanism. Results are in excellent agreement for all the chemical species and the temperature. Note that, although one- and two-dimensional SIM's are able to recover the most of the dynamics of major species and of the temperature, the minority species (such as radicals ${\rm HO}_2$ and ${\rm H}_2{\rm O}_2$) do require high dimensional manifolds ($q \ge 3$) to be correctly predicted.

For the sake of clarity, we outline below the steps leading to the computation of a $q$-dimensional closure corresponding to a macroscopic state $\xi = \left( \xi^1,...,\xi^q\right)$, by the local RRM for the above kinetic system:
\begin{enumerate}
  \item Set up the initial SIM dimension (e.g., $q=1$);
	\item Set up a convergence criterion (\ref{threshold}), and the maximal number of iterations $N$;
	\item Compute the initial coordinates of the pivot $\psi^*$, which amounts to solving a non linear algebraic system, $\nabla \bar G=0$, e.g. by Newton-Raphson iterations;
	\item Compute $q$ secondary nodes $\psi^k$ by the linear algebraic system (\ref{QEG.sys}); 
	\item Update the coordinates of both the pivot and secondary nodes by the RRM equation (\ref{RRpivot});
	\item Check convergence;
	\item if no convergence is achieved after $N$ iterations, then update the SIM dimension $q=q+1$ and go to 3;
	\item exit.
\end{enumerate}

The above illustration demonstrates that the suggested RRM method is able to accurately recover the dynamics of a complex system. Moreover, here we adopted the automatic criterion (\ref{threshold}) to choose the number of reduced degrees of freedom (macroscopic important variables), which are {\em strictly} needed to reproduce the phenomenon under study. The latter features make the RRM, on one side, a pretty useful tool for the efficient computation of large dissipative systems. Most importantly, on the other side, it enables to gain a better physical understanding about a complex phenomena by addressing the issue of its minimal description.


%
%
\section{Conclusion}\label{conclusion.sec}
%
To conclude, we addressed here the fundamental problem of the minimal description of a complex dissipative system which is a challenging issue in Physics. Our approach is based on a simulation (instead of a solution) of the fundamental film equation of dynamics (\ref{film.dynamics}). 
We stress that it is the RRM realization which is able to unfold the full power of the Method of Invariant Manifold which was not possible before, such as the adaptive construction of high dimensional manifolds (i.e. $q \ge 3$, with $q$ varying from a region of the phase-space to another). On the practical side, RRM is pretty simple as it is based on a direct integration of the film equation plus redistribution. The key point realized in this paper is that the latter simulates subtraction of slow motion from the film dynamics, the step which is hard to control in more conventional approaches to the film equation \cite{GKZ04}. In that respect, the RRM is similar in its spirit (but certainly not in the implementation) to other successful simulation strategies such as the Direct Simulation Monte Carlo method \cite{Bird} which replaces the solution of the Boltzmann equation by a stochastic simulation of "collisions". 
We stress that, suitable macroscopic variables depend on the specific phenomenon (e.g. velocity moments of the distribution function for describing gas kinetics \cite{Grad49,MullerRuggeribook}). The methodology developed in this paper addresses the general problem of minimal macroscopic description by letting the system decide how many {\em important} variables are to be considered.

Examples presented above convincingly show that RRM achieves all the objectives set for obtaining the accurate reduced description, whereas the resulting adaptively reduced models reveal new physical knowledge of a complex dissipative system (i.e. its minimal description), and can be used for its computationally efficient simulation. We should stress that fully adaptive construction of heterogeneous slow invariant manifolds as in the case of hydrogen-air mixture is difficult if at all possible with any other model reduction technique \cite{ModRedCollection}. Finally, while we focused on the important class of dissipative systems arising in combustion, we look forward to generalization of the above technique of simplification to other dissipative systems such as master and Fokker-Planck equations and  other complex dynamics.

%

\appendix

\section{Detailed reaction for hydrogen and air}\label{append01}
In the table \ref{table.mech}, we report the list of all reaction steps involved in the combustion mechanism for hydrogen and air adopted in section \ref{illustration.h2o2}, where $n=9$ species ($H_2$, $N_2$, $H$, $O$, $OH$, $O_2$, $H_2O$, $HO_2$, $H_2O_2$) and $d=3$ elements ($H$, $O$, $N$) are involved in $r=21$ elementary reversible steps. The system of kinetic equations is formulated according to the (\ref{mol.concentration.eq}) and (\ref{mass.action.law}), where the reaction constant $k^+_s$ of the $s$-th step is determined by the Arrhenius law (\ref{arrhenius}) with the coefficients $A_s$, $n_s$ and $Ea_s$ from table \ref{table.mech}. In the following, the symbol $M$ represents an additional species, whose concentration $c_M$ denotes a weighted sum of the concentration of all species (third-body reaction):
\begin{equation}
c_M=\sum_{i=1}^n a_i c_i,
\end{equation}
$a_i$ being the {\em third-body efficiencies}. In the reactions N. 5, 6, 7 ,8, it is adopted $a_{H_2O}=11.0$, $a_{H_2}=1.5$, and $a_i=1$ for all other species. Finally, the steps N. 9 and 16 are typical {\em fall-off} reactions, where the reaction constant $k^+_s$ remarkably depends on the mixture pressure. In this case, $k_\infty^+$ and $k_0^+$ are the reaction constants in the high- and low-pressure limit, respectively, and the reaction constant reads:
\begin{equation}
k_s^+= k_\infty ^+ F P_r/\left( 1+ P_r\right), 
\end{equation}
with $P_r=k_0^+ c_M / k_\infty^+$, and $F$ given by the Troe function (see \cite{Lawbook} for the details). In particular, in the reaction step N. 9 the third-body efficiency are $a_{H_2O}=10$, $a_{O_2}=-0.22$, in the reaction step N. 16 $a_{H_2O}=11$, $a_{H_2}=1.5$, whereas in both cases $a_i=1$ for the rest of the species.  
\begin{table}[htbp]
\centering
\begin{tabular}{l|c|c|c}
\hline
\hline
\multicolumn{1}{l|}{\textbf{Reaction}} &
\multicolumn{1}{c|}{$A_s$} &
\multicolumn{1}{c|}{$n_s$} &
\multicolumn{1}{c}{$Ea_s$} \\
\hline
1. $H+O_2 \rightleftharpoons O+OH$  &	  $3.55 \times 10^{15}$ 				& -0.41 & 16.6  \\
2. $O+H_2 \rightleftharpoons H+OH$    &	$5.08 \times 10^{4}$ 					& 2.67 	& 6.29  \\
3. $H_2+OH \rightleftharpoons H_2O+H$ & $2.16 \times 10^{8}$ 					& 1.51 	& 3.43  \\
4. $O+H_2O \rightleftharpoons OH+OH$  &	$2.97 \times 10^{6}$					&	2.02	&	13.4	\\
5. $H_2+M \rightleftharpoons H+H+M$		&	$4.58 \times 10^{19}$					&	-1.40	&	104.38\\
6. $O+O+M \rightleftharpoons O_2+M$		&	$6.16 \times 10^{15}$					&	-0.50	&	0.00  \\
7. $O+H+M \rightleftharpoons OH+M$		&	$4.71 \times 10^{18}$					&	-1.0	&	0.00	\\
8. $H+OH+M \rightleftharpoons H_2O+M$	&	$3.8 \times 10^{22}$					&	-2.00	&	0.00	\\
9. $H+O_2(+M) \rightleftharpoons HO_2(+M)^a \quad k_0^+$	&	$6.37 \times 10^{20}$	&	-1.72	&	0.52	\\
   $\quad \quad \quad \quad \quad \quad \quad \quad \quad \quad \quad \quad \quad \quad \quad \; k_{\infty}^+$	&	$1.48 \times 10^{12}$	&	0.60	&	0.00	\\
10.$HO_2+H \rightleftharpoons H_2+O_2$				&	$1.66 \times 10^{13}$	&	0.00	&	0.82	\\
11.$HO_2+H \rightleftharpoons OH+OH$					&	$7.08 \times 10^{13}$	&	0.00	&	0.30	\\
12.$HO_2+O \rightleftharpoons O_2+OH$					&	$3.25 \times 10^{13}$	&	0.00	&	0.00	\\
13.$HO_2+OH \rightleftharpoons H_2O+O_2$			&	$2.89 \times 10^{13}$	&	0.00	&	-0.50	\\
14.$HO_2+HO_2 \rightleftharpoons H_2O_2+O_2$	&	$4.20 \times 10^{14}$	&	0.00	&	11.98	\\
15.$HO_2+HO_2 \rightleftharpoons H_2O_2+O_2$	&	$1.30 \times 10^{11}$	&	0.00	&	-1.63	\\
16.$H_2O_2(+M) \rightleftharpoons 2OH(+M)^b \quad \quad k_0^+$	&	$1.20 \times 10^{17}$	&	0.00	&	45.5	\\
	 $\quad \quad \quad \quad \quad \quad \quad \quad \quad \quad \quad \quad \quad \quad \quad \; k_{\infty}^+$	&	$2.95 \times 10^{14}$	&	0.00	&	48.4	\\				
17.$H_2O_2+H \rightleftharpoons H_2O+OH$			&	$2.41 \times 10^{13}$	&	0.00 	&	3.97 	\\
18.$H_2O_2+H \rightleftharpoons HO_2+H_2$			&	$4.82 \times 10^{13}$	&	0.00  &	7.95	\\
19.$H_2O_2+O \rightleftharpoons OH+HO_2$			&	$9.55 \times 10^{6}	$	&	2.00  &	3.97	\\
20.$H_2O_2+OH \rightleftharpoons HO_2+H_2O$		&	$1.00 \times 10^{12}$	&	0.00	&	0.00	\\
21.$H_2O_2+OH \rightleftharpoons HO_2+H_2O$		&	$5.8 \times 10^{14}	$ &	0.00	&	9.56	\\
\hline
\end{tabular} 
\caption{Detailed $H_2$-air reaction mechanism. Units are $cm^3$, $mol$, $sec$, $Kcal$ and $K$. $^a$Troe parameter is: $0.8$. $^b$Troe parameter is: $0.5$.}\label{table.mech}
\end{table}
\end{document}